\documentstyle[12pt]{article}
\input html.sty
\begin{document}
\title{MATTERS OF GRAVITY, The newsletter of the APS TIG on Gravitation}
\begin{center}
{ \Large {\bf MATTERS OF GRAVITY}}\\
\bigskip
\hrule
\medskip
{The newsletter of the Topical Group in Gravitation of the American Physical 
Society}\\
\medskip
{\bf Number 9 \hfill Spring 1997}
\end{center}
\begin{flushleft}

\tableofcontents
\vfill
\section*{\noindent  Editor\hfill}
%{\bf Editor:}

\medskip
Jorge Pullin\\
\smallskip
Center for Gravitational Physics and Geometry\\
The Pennsylvania State University\\
University Park, PA 16802-6300\\
Fax: (814)863-9608\\
Phone (814)863-9597\\
Internet: 
\htmladdnormallink{\protect {\tt pullin@phys.psu.edu}}
{mailto:pullin@phys.psu.edu}\\
WWW: \htmladdnormallink{\protect {\tt http://www.phys.psu.edu/PULLIN}}
{http://www.phys.psu.edu/PULLIN}
\begin{rawhtml}
<P>
<BR><HR><P>
\end{rawhtml}
%{\bf \Large Contents:}
\end{flushleft}
\pagebreak
\section*{Editorial}

Nothing profound to say in this editorial, just to thank the
contributors and correspondents that make this newsletter possible and
as usual to remind everyone that suggestions for authors/topics for
the newsletter are very welcome.

The next newsletter is due September 1st.  If everything goes well this
newsletter should be available in the gr-qc Los Alamos archives under
number gr-qc/9702010. To retrieve it send email to 
\htmladdnormallink{gr-qc@xxx.lanl.gov}{mailto:gr-qc@xxx.lanl.gov}
(or 
\htmladdnormallink{gr-qc@babbage.sissa.it}{mailto:gr-qc@babbage.sissa.it} 
in Europe) with Subject: get 9702010
(numbers 2-8 are also available in gr-qc). All issues are available in the
WWW:\\\htmladdnormallink{\protect {\tt
http://vishnu.nirvana.phys.psu.edu/mog.html}}
{http://vishnu.nirvana.phys.psu.edu/mog.html}\\ 
A hardcopy of the newsletter is
distributed free of charge to the members of the APS
Topical Group on Gravitation. It is considered a lack of etiquette to
ask me to mail you hard copies of the newsletter unless you have
exhausted all your resources to get your copy otherwise.

If you have comments/questions/complaints about the newsletter email
me. Have fun.
\bigbreak

\hfill Jorge Pullin\vspace{-0.8cm}
\section*{Correspondents}
\begin{itemize}
\item John Friedman and Kip Thorne: Relativistic Astrophysics,
\item Raymond Laflamme: Quantum Cosmology and Related Topics
\item Gary Horowitz: Interface with Mathematical High Energy Physics and
String Theory
\item Richard Isaacson: News from NSF
\item Richard Matzner: Numerical Relativity
\item Abhay Ashtekar and Ted Newman: Mathematical Relativity
\item Bernie Schutz: News From Europe
\item Lee Smolin: Quantum Gravity
\item Cliff Will: Confrontation of Theory with Experiment
\item Peter Bender: Space Experiments
\item Riley Newman: Laboratory Experiments
\item Warren Johnson: Resonant Mass Gravitational Wave Detectors
\item Stan Whitcomb: LIGO Project
\end{itemize}
\vfill
\pagebreak

\section*{\centerline {April 1997 Joint APS/AAPT Meeting}}
\addtocontents{toc}{\protect\medskip}
\addtocontents{toc}{\bf News:}
\addtocontents{toc}{\protect\medskip}
\addcontentsline{toc}{subsubsection}{\it  April 1997 Joint APS/AAPT Meeting,
by Beverly Berger}
\begin{center}
    Beverly Berger\\
               Oakland University\\
\htmladdnormallink{berger@vela.oakland.edu}
{mailto:berger@vela.oakland.edu}
\end{center}
\parindent=0pt
\parskip=5pt

The 1997 Joint APS/AAPT Meeting will be held 18-21 April 1997 in
Washington, DC and features, among other things, the activities of the
Topical Group in Gravitation (GTG) including invited and focus
sessions sponsored by GTG, as well as the GTG annual business
meeting. The Ligo Research Community will also hold its meeting here.

\medskip
Invited session speakers and talk titles:
\medskip

{\em Sources and Detection of Gravitational Radiation} (sponsored by the
Topical Group in Gravitation):

$\bullet$     Jorge Pullin, Penn State,\hfil\break
     ``Analytic insights into the collision of two black holes"

$\bullet$      Joan Centrella, Drexel,\hfil\break
     ``Gravitational Radiation from Inspiralling Binary Neutron Stars 

$\bullet$      Bruce Allen, UMW and LIGO,\hfil\break
     ``Will LIGO Detect a Gravitational Wave Background?''

$\bullet$      David H. Shoemaker, MIT LIGO,\hfil\break
     ``LIGO: Observatory and Instrument Status''
\medskip

{\em Frontiers in Theoretical Physics} (sponsored by the Topical Group in
Gravitation and the Division of Particles and Fields)

$\bullet$      Robert M. Wald, Chicago,\hfil\break
     ``Gravitational Collapse and Cosmic Censorship''

$\bullet$      Abhay Ashtekar, Penn State,\hfil\break
     ``Large Quantum Gravity Effects in Unexpected Domains''

$\bullet$      Juan Maldacena, Rutgers,\hfil\break
     ``Black Holes and String Theory''
\medskip

{\em Sensitive Mechanical Measurements and the Detection of Gravitational Waves}
(sponsored by the Topical Group in Gravitation and the Topical Group in
Fundamental Constants and Precision Measurement)

$\bullet$      Peter Saulson, Syracuse,\hfil\break
     ``Internal Friction, Brownian Motion, and General Relativity: Low Noise
Mechanics and the Challenge of Gravitational Wave Detection''

$\bullet$      W.O. Hamilton, LSU,\hfil\break
`` Resonant Gravity Wave Detectors---What They
     Have Done and What They Can Do''

$\bullet$      Jennifer Logan, Caltech/LIGO,\hfil\break
`` Noise Behavior of the LIGO 40m
     Interferometer for Gravitational Wave Detection''

$\bullet$      Mark Bocko, Rochester,\hfil\break
`` Weak Force Detection Strategies, Where
Quantum Mechanics and Gravitation Really Meet!''

\medskip
The GTG has organized two focus sessions with invited and contributed
talks on specific topics of interest to the GTG membership. The first,
chaired by Leonard Parker, is on "Black Hole Evaporation, Formation,
and Entropy" while the second, chaired by Bill Hamilton, is on
"Analyzing Data from Gravitational-Wave Detectors."

The GTG will also organize contributed sessions on other areas of
experimental and theoretical gravitational physics.

More details are available on the APS Meetings
(
\htmladdnormallink{\protect {\tt http://www.aps.org/meet/meetcal.html}}
{http://www.aps.org/meet/meetcal.html})
web page.  

The deadline for contributed abstracts has passed.

\vfill
\pagebreak

\section*{\centerline {TGG News}}
\addtocontents{toc}{\protect\medskip}
\addcontentsline{toc}{subsubsection}{\it  TGG News, by Jim Isenberg}
\begin{center}
    Jim Isenberg, Sec/Treas of the TGGravitation\\
               University of Oregon\\
\htmladdnormallink{jim@newton.uoregon.edu}
{mailto:jim@newton.uoregon.edu}
\end{center}

$\bullet$ {\em Election News}

We completed our second election this past November. The results were  
extremely close! (3 votes out of about 120 determined two of the  
races) The winning candidates for each contested office are as  
follows: 

Vice Chair: Rainier Weiss
Executive Committee (Slot 1): Sam Finn
Executive Committee (Slot 2): Mac Keiser
These people will take office just before the spring meeting, in  
April.

$\bullet$ {\em  Press Agent}

The APS is encouraging each unit (ie, each topical group, each  
division, etc) to set up a procedure for getting interesting news  
items out to the press.  I think that the simplest thing (at least  
for those of us not used to doing this sort of thing on their own) is  
for members to send such items to the Sec/Treas (currently me), and  
then I will forward them with comments to the APS/AIP people whose  
job it is to publicize physics and to work with the press.  Please  
don't be shy about sending me such items. My e-mail is  
\htmladdnormallink{jim@newton.uoregon.edu}
{mailto:jim@newton.uoregon.edu}, and my fax is  (541)346-5217.

$\bullet$ {\em  Centenary  Speaker List}

The APS Centenary is coming up in 1999. To help celebrate, the APS is  
planning to set up a list of top notch speakers who will be called  
upon to give special colloquia and special public lectures around the  
country. If you wish to volunteer yourself or anyone else for  
inclusion on this list, please let me know.

$\bullet$ {\em  Centenary Exhibits}

A large part of the Centenary celebration will focus on the APS  
meeting held in Atlanta in March of 1999.  The APS is soliciting  
exhibits and displays for this meeting. They would like to have a lot  
of them. If any of you have any such things to offer, again, please  
let me know.

$\bullet$ {\em  The Malcolm MacCallum GR Notices}

Malcolm MacCallum very generously maintains an electronic GR news  
service which comes out monthly. While it is likely that most of us  
subscribe, it may be that some of you do not. If you do not, but  
would like to, just send an e-mail with your name and e-mail address  
and phone and fax to 
\htmladdnormallink{M.A.H.MacCallum@qmw.ac.uk}
{mailto:M.A.H.MacCallum@qmw.ac.uk}

We all greatly appreciate this generous service.

$\bullet$ {\em Changing  Your Address}

If you are changing your address--electronic or postal--the APS folks  
(including us at the Topical  Group) would like to know. Just send a  
note to 
\htmladdnormallink{membership@aps.org}
{mailto:membership@aps.org}
 with the relevant information.

\parindent=10pt
\parskip=5pt
\vfill
\pagebreak
\section*{\centerline {Report from NSF}}
\addtocontents{toc}{\protect\medskip}
\addcontentsline{toc}{subsubsection}{\it Report from NSF, by David Berley}
\begin{center}
                     David Berley\\
               NSF Gravitational Physics\\
\htmladdnormallink{db@einstein.mps.nsf.gov}
{mailto:db@einstein.mps.nsf.gov}\\
\end{center}

The LIGO project is nearing completion. In mid 1999 two
interferometers are scheduled to be installed at the Hanford,
Washington site and shortly thereafter a third is to be installed at
the Livingston, Louisiana site.  By the end of 2001 all three
interferometers are planned to operate in coincidence with a strain
sensitivity of $10^{-21}$. In anticipation of the operations phase of
LIGO, NSF has been encouraging the scientific community to participate
in the use and development of LIGO. 

On June 25, 1996, NSF convened the Panel on the Long Range Use of LIGO
(henceforth to be called the Panel).  The Panel was asked to advise
NSF on the policies, procedures and resources required to stimulate
and support outstanding investigations at LIGO.  The Panel was also
asked to comment explicitly on the respective roles of the LIGO
Project and the NSF in the organization, review, and funding of the
scientific observations and the detector R\&D.  Thirdly, the Panel was
asked to estimate the size of the users community and to advise on the
funds that would be required over the next decade for LIGO to be an
effective user facility.  The Panel members were drawn from several
areas of the physical sciences including, of course, gravitational
physics, but also including disciplines requiring large facilities.
Appearing before the Panel were representatives of groups planning to
use LIGO, heads of gravitational wave facilities abroad, and the Chair
of the LIGO Research Community. 

The Panel made recommendations on organization, procedures, policies,
data handling and funding.  The Panel's report covers a broad range of
issues and includes 13 recommendations.  I will discuss only a few of
these here.  The complete report is available on the Web at
\htmladdnormallink{http://www.nsf.gov:80/mps/phy/ligorpt.htm} 
{http://www.nsf.gov:80/mps/phy/ligorpt.htm}. 

The Panel began with the premise that the essential near-term priority
for LIGO is to achieve a definitive detection of gravitational waves.
This achievement will validate the system and inaugurate the field of
gravitational wave astronomy.  Despite the impressive achievements for
the initial LIGO detector, greater sensitivity may still be required
for the unambiguous detection of astronomical sources.  Even if
signals are detected with the initial detector, a greater signal to
noise ratio will be required to pursue gravitational wave astronomy.
Therefore, in the medium term accompanying and following the
commissioning, the LIGO Project must be focused on an aggressive R\&D
effort to achieve an improvement of 10 to 100 times greater strain
sensitivity.  The Panel noted that, as part of this effort, it is
imperative to develop extended national and international
collaborations in order to bring the best possible people, ideas and
technologies to bear on the detection of gravitational wave radiation.
The Panel recommended that these collaborations be integrated with
centralized R\&D coordination and LIGO Project management.  The
collaborations developed during this intermediate term will form the
basis of the community that will carry LIGO into the longer term
marked by the transition of gravitational wave detection into an
observational science. 

The Panel recommended that the LIGO Project should evolve from the
current single management into two distinct entities: 1) a LIGO
Laboratory and 2) a formally organized initial Collaboration.  The
Laboratory will provide the infrastructure that makes it possible for
instruments to be developed, built and operated.  The Collaboration
will build, commission and exploit the initial detector and develop
improvements to enhance its sensitivity.  The Panel recommended that
the Collaboration devise its own plan for internal governance,
including clear procedures for the admission of new members.  It will
be essential for the Collaboration to have a spokesperson who will
communicate with the outside community. 

The Panel recommended the formation of a Program Advisory Committee
(PAC) to provide advice on the formation of the Collaboration, the
acceptance of other collaborators, the selection of R\&D projects and
the assignment of priorities.  The Panel proposed that the PAC be part
of the NSF review process for LIGO related proposals. 

Following this advice, the NSF recently asked the LIGO Project for its
opinion on several pending LIGO-related proposals. In formulating the
Project opinion, Professor Barish, head of the LIGO Project, obtained
an internal staff review of the proposals and then asked the PAC also
to review them.  With this action, the PAC review became part of the
NSF review process. 

All of these pending proposals will also be processed through the
normal NSF peer review procedure.  The evaluations by the LIGO staff
and the PAC will be forwarded to the NSF reviewers.  This augmentation
of the NSF review procedure appears to be working well and, perhaps
with some minor modification, will become the standard for processing
LIGO-related proposals. 

The Panel on the Use of LIGO recognized that LIGO is a national
facility and has an obligation to make its data available to those who
can make effective use of them.  The Panel also recognized that the
detection of gravitational radiation will be of such monumental
importance that we can ill afford the report of a false signal. 

Therefore, the Panel suggested that community involvement in data
handling during the initial stage be achieved by drawing into the
Collaboration those who would work directly with the in-house LIGO
team in developing data analysis tools.  During this stage, data will
be given only to the Collaboration and publications will require the
approval of the Collaboration and in some cases the LIGO Principal
Investigator.  The Panel noted that the possibility of developing
distributed data products should be considered after gravitational
radiation events have been detected and the detectors are well
understood. 

The Foundation has encouraged the collaboration of outside scientists
with the LIGO project first to contribute to the initial detector and
second, to address the R\&D required to improve its sensitivity.  The
effective use of NSF resources will involve collaborations with the
LIGO Project and later with the Collaboration.  These arrangements are
being codified through substantive memoranda of understanding (MOUs)
among the parties involved.  These MOUs are the instruments through
which NSF will know that each party understands and agrees to its role
and scope of activity and the role of its collaborators. 

The development of advanced detectors, those beyond the initial LIGO
detector and its enhancements, will also require a coherent effort.
Plans for handling such proposals await the formation of the
Collaboration and the establishment of proper linkages with the
Collaboration and the LIGO Laboratory.

\vfill\pagebreak
\section*{\centerline {We hear that...}}
\addtocontents{toc}{\protect\medskip}
\addcontentsline{toc}{subsubsection}{\it We hear that... by Jorge Pullin}
\begin{center}
                     Jorge Pullin\\
               Penn State\\
\htmladdnormallink{pullin@phys.psu.edu}
{mailto:pullin@phys.psu.edu}\\
\end{center}

\parindent=0pt
\parskip=10pt

\bigskip

Several members of the Topical Group on Gravitation were elected APS 
fellows.

\bigskip

{\em Nominated through the Topical Group on Gravitation:}

$\bullet$ Robert Wald, University of Chicago, ``For his contributions
to the understanding of classical and quantum gravity; especially for
his seminal role in the development of a rigorous basis for quantum
field theory in curved spacetime.''

$\bullet$ Rainer Weiss, MIT, ``For his pioneering work in the
development of laser-interferometric detectors for gravitational
radiation, and his contributions to the study of the spectrum and
anisotropy of the cosmic microwave background. ``

\bigskip

{\em Nominated through the Division of Astrophysics:}

$\bullet$ James Bardeen, University of Washington, ``For his seminal
contributions to the theory of cosmological density perturbations,
relativistic astrophysics, and galactic structure.''

$\bullet$ Edmund Bertschinger, MIT, ``For his outstanding
contributions to theoretical cosmology, especially in the
understanding of structure formation in the universe.''

\bigskip

{\em Nominated through the Forum for International Physics:}

$\bullet$ Rodolfo Gambini, Universidad de la Republica, Montevideo,
Uruguay, ``For distinguished research in field theory and gravitation,
notably on geometrical techniques and the loop representation of gauge
theories, and for mentoring theoretical physicists in Latin America.''

\bigskip

The complete list of new fellows can be seen in 
\htmladdnormallink{http://www.aps.org/fellowship/96indx.html}
{http://www.aps.org/fellowship/96indx.html}

\parindent=10pt
\parskip=5pt

\vfill
\pagebreak
\section*{\centerline {General relativity in the global positioning system}}
\addtocontents{toc}{\protect\bigskip}
\addtocontents{toc}{\bf Research briefs:}
\addtocontents{toc}{\protect\medskip}
\addcontentsline{toc}{subsubsection}{\it 
GR in GPS, by Neil Ashby}
\begin{center}
                     Neil Ashby\\
               University of Colorado\\
\htmladdnormallink{n\_ashby@mobek.colorado.edu}
{mailto:n\_ashby@mobek.colorado.edu}
\end{center}

	The Global Position System (GPS) consists of 24 earth-orbiting
satellites, each carrying accurate, stable atomic clocks.  Four
satellites are in each of six different orbital planes, of inclination
55 degrees with respect to earth's equator.  Orbital periods are 12
hours (sidereal), so that the apparent position of a satellite against
the background of stars repeats in 12 hours.  Clock-driven
transmitters send out synchronous time signals, tagged with the
position and time of the transmission event, so that a receiver near
the earth can determine its position and time by decoding navigation
messages from four satellites to find the transmission event
coordinates, and then solving four simultaneous one-way signal
propagation equations.  Conversely, gamma-ray detectors on the
satellites could determine the space-time coordinates of a nuclear
event by measuring signal arrival times and solving four one-way
propagation delay equations.

	Apart possibly from high-energy accelerators, there are no
other engineering systems in existence today in which both special and
general relativity have so many applications.  The system is based on
the principle of the constancy of c in a local inertial frame: the
Earth-Centered Inertial or ECI frame.  Time dilation of moving clocks
is significant for clocks in the satellites as well as clocks at rest
on earth.  The weak principle of equivalence finds expression in the
presence of several sources of large gravitational frequency shifts.
Also, because the earth and its satellites are in free fall,
gravitational frequency shifts arising from the tidal potentials of
the moon and sun are only a few parts in $10^{16}$ and can be neglected.

	The Sagnac effect has an important influence on the
system. Since most GPS users are at rest or nearly so on earth's
surface, it would be highly desirable to synchronize clocks in a
rotating frame fixed to the earth (an Earth-Fixed, Earth-Centered
Frame or ECEF Frame).  However because the earth rotates, this is
prevented by the Sagnac effect, which is large enough in the GPS to be
significant.  Inconsistencies occurring in synchronization processes
conducted on the Earth's surface by using light signals, or with
slowly moving portable clocks, are path-dependent and can be many
dozens of nanoseconds, too large to tolerate in the GPS.  Thus the
Sagnac effect forces a different choice for synchronization
convention.  Also, the path of a signal in the ECEF is not "straight."
In the GPS, synchronization is performed in the ECI frame; this solves
the problem of path-dependent inconsistencies.

	Several sources of relativistic effects enter in determining
the unit of time, the SI second as realized by the U. S. Naval
Observatory (USNO).  For a clock fixed on earth, time dilation arising
from earth's spinning motion can be viewed alternatively as a
contribution, in the ECEF frame, to the total effective gravitational
potential which also includes contributions arising from earth's
non-sphericity.  Earth-fixed clocks placed on the same equipotential
surface of this effective field all beat at the same rate.  Over the
span of geological time, the earth's figure has distorted so that it
nearly matches one of these gravitational equipotentials--the earth's
geoid at mean sea level.  The SI second is defined by the rate of
atomic clocks on the geoid.  This rate is determined to sufficient
accuracy, relative to clocks at infinity, by three effects: time
dilation due to earth's spin, and frequency shifts due to the monopole
and quadrupole potentials of earth.

	In General Relativity (GR), coordinate time, such as is
expressed approximately by a slow-motion, weak-field metric, covers
the solar system.  The proper time elapsed on a moving clock depends
on the clock's position and velocity in the fields of nearby masses,
and can be computed in terms of the elapsed coordinate time if the
velocities, positions, and masses are known.  Conversely, the elapsed
coordinate time can be computed by integrating corrections to the
proper time.

	The concept of coordinate time in a local inertial frame is
established for the GPS as follows.  In the local ECI frame, imagine a
network of atomic clocks at rest and synchronized using constancy of
c.  To each real, moving clock apply corrections to yield a paper
clock which then agrees with one of these hypothetical clocks in the
underlying inertial frame, with which the moving clock instantaneously
coincides.  The time resulting from such corrections is then a
coordinate time, free from inconsistencies, whose rate is determined
by clocks at rest on the earth's rotating geoid.

	Relativistic effects on satellite clocks can be combined in
such a way that only two corrections need be considered.  First, the
average frequency shift of clocks in orbit is corrected downward in
frequency by $446.47$ parts in $10^{12}$.  This is a combination of five
different sources of relativistic effects: gravitational frequency
shifts of ground clocks due to earth's monopole and quadrupole
moments, gravitational frequency shifts of the satellite clock, and
second-order Doppler shifts from motion of satellite and earth-fixed
clocks.  Second, if the orbit is eccentric, an additional correction
arises from a combination of varying gravitational and motional
frequency shifts as the satellite's distance from earth varies.  This
correction is periodic and is proportional to the orbit eccentricity.
For an eccentricity of $.01$, the amplitude of this term is $23$ ns.  Due
to a shortage of computer resources on satellites in the early days of
GPS, it was decided that this latter correction was to be the
responsibility of software in GPS receivers.  It is a correction which
must be applied to the broadcast time of signal transmission, to
obtain the coordinate time epoch of the transmission event in the ECI
frame.

	At the time of launch of the first NTS-2 satellite (June
1977), which contained the first Cesium clock to be placed in orbit,
there were some who doubted that relativistic effects were real.  A
frequency synthesizer was built into the satellite clock system so
that after launch, if in fact the rate of the clock in its final orbit
was that predicted by GR, then the synthesizer could be turned on
bringing the clock to the coordinate rate necessary for operation.
The atomic clock was first operated for about 20 days to measure its
clock rate before turning on the synthesizer.  The frequency measured
during that interval was $+442.5$ parts in $10^{12}$ faster than clocks on
the ground; if left uncorrected this would have resulted in timing
errors of about $38,000$ nanoseconds per day.  The difference between
predicted and measured values of the frequency shift was only $3.97$
parts in $10^{12}$, well, within the accuracy capabilities of the orbiting
clock.  This then gave about a $1\%$ validation of the combined motional
and gravitational shifts for a clock at $4.2$ earth radii.

	At present one cannot easily perform tests of relativity with
the system because the SV clocks are actively steered to be within 1
microsecond of Universal Coordinated Time (USNO).

	Several relativistic effects are too small to affect the
system at current accuracy levels, but may become important as the
system is improved; these include gravitational time delays, frequency
shifts of clocks in satellites due to earth's quadrupole potential,
and space curvature.

	This system was intended primarily for navigation by military
users having access to encrypted satellite transmissions which are not
available to civilian users.  Uncertainty of position determination in
real time by using the Precise Positioning code is now about $2.4$
meters.  Averaging over time and over many satellites reduces this
uncertainty to the point where some users are currently interested in
modelling many effects down to the millimeter level.  Even without
this impetus, the GPS provides a rich source of examples for the
applications of the concepts of relativity.

	New and surprising applications of position determination and
time transfer based on GPS are continually being invented.  Civilian
applications include for example, tracking elephants in Africa,
studies of crustal plate movements, surveying, mapping, exploration,
salvage in the open ocean, vehicle fleet tracking, search and rescue,
power line fault location, and synchronization of telecommunications
nodes.  About 60 manufacturers now produce over 350 different
commercial GPS products.  Millions of receivers are being made each
year; prices of receivers at local hardware stores start in the
neighborhood of \$200.

\vfill
\pagebreak
\section*{\centerline {What Happens Near the Innermost Stable Circular Orbit?}}
\addtocontents{toc}{\protect\medskip}
\addcontentsline{toc}{subsubsection}{\it What Happens Near the Innermost Stable
Circular Orbit? by Doug Eardley}
\begin{center}
Doug Eardley, ITP, UCSB\\
\htmladdnormallink{doug@itp.ucsb.edu}{mailto:doug@itp.ucsb.edu}
\end{center}

In decaying binary systems consisting of neutron stars or black holes, a
lot can happen near the {\em innermost stable circular orbit,} or {\em
ISCO}, as a number of theorists have shown over the last couple of years.
A number of questions still remain open though:  Where the ISCO is
located, what physics determines it, and what observable effects ensue.
First of all, the ISCO is not even precisely defined.

For the special case of test particle orbiting
around a black hole (or sufficiently compact neutron star) the ISCO is
well defined, as we all learned in our relativity courses: at $r=6M$ for
a non-rotating black hole, or $r=M$ for a maximally rotating black hole.
But the binary orbit of two massive stars evolves due to gravitational
radiation reaction, and the ISCO is really the fuzzily-defined radius
where slow orbital shrinkage goes over to rapid plunge.  Because
energy loss by gravitational waves is a bit inefficient even under
the best of circumstances, the orbital shrinkage is probably slow
enough that the ISCO is reasonably well defined in practice:  For
instance, if a distant gravity wave observer plots plots gravity wave
amplitude as a function of time, a distinct feature will appear.
The ``ISCO radius" is a phrase commonly used, but must be deprecated
because there is no unique (or at least generally agreed upon) way to
measure it invariantly.

Sharpening the issue further is its relevance to the Grand Challenge, as was
\htmladdnormallink
{reported}{http://vishnu.nirvana.phys.psu.edu/mog/mog8/node13.html#SECTION000130000000000000000}
by Sam Finn in \htmladdnormallink
{MOG8}{http://vishnu.nirvana.phys.psu.edu/mog/mog8/mog8.html}\@.  The
first 3D codes able to evolve such a binary must begin near the
ISCO\@, so the numerical relativists need to know where it is, and how
to model the system for initial conditions near it.  The gravity
waveforms observable in LIGO, VIRGO and LISA will depend on this issue
(Flanagan \& Hughes, \htmladdnormallink
{gr-qc/9701039}{http://xxx.lanl.gov/abs/gr-qc/9701039}), and so may
gamma-ray bursts (Rees, \htmladdnormallink
{astro-ph/9701162}{http://xxx.lanl.gov/abs/astro-ph/9701162}).

The purpose of this report is to give a quick and nontechnical introduction
to these rapidly evolving issues, along with links to the literature for
those who want to delve into the whole story.  I have probably missed
some references; if so, let me know.

The earliest approach to the ISCO is the post-Newtonian one, which has
gone through a long development, most notably by Will and collaborators.
Two recent papers which reference this body of work are Lai \& Wiseman
(\htmladdnormallink
{gr-qc/9609014}{http://xxx.lanl.gov/abs/gr-qc/9609014}),
and Will \& Wiseman ({\em Phys.\ Rev.\ }{\bf D54}, p.4813, 1996; 
\htmladdnormallink{gr-qc/9608012}{http://xxx.lanl.gov/abs/gr-qc/9608012}).
If one ignores radiation, the ISCO is well defined here, and if
radiation reaction is included, the ISCO is indeed ``reasonably well defined"
({\em Phys.\ Rev.\ }{\bf D47}, p.3281, 1993).
Sensible answers emerge, but the calculations must be carried to high order,
and theoretically speaking it not clear they converge.  For binaries
involving neutron stars, an important physical effect is the tidal
interaction between the two stars, which may disrupt the stars (as pointed
out in Newtonian theory by Dai, Rasio and Shapiro, {\em Ap.J.\ }{\bf 420},
p.811, 1994) before the ISCO is reached, and which also can affect the
location of the ISCO.

More recently, Taniguchi and Nakamura
(\htmladdnormallink{astro-ph/9609009}{http://xxx.lanl.gov/abs/astro-ph/9609009})
have developed an analytic method based on fluid ellipsoids and
a pseudo-relativistic form of the Newtonian gravitational potential,
of the form $GM/(r-r_{pseudo})$.  This potential doesn't
enjoy a clear justification from GR theory (beyond a reasonably
accurate reproduction of test particle orbits) but at least it has
the virtue of simplicity.  They are able to do a parameter study
of tidal effects as a function of orbital compactness, and they clearly
show in their models the dividing line between cases where tidal effects
dominate in creating the ISCO, and cases where relativistic orbital
effects dominate.

Turning to full GR theory,
one way to make the ISCO well defined is to put our binary into a
large cavity with walls made of a mythical material that perfectly reflects
gravity waves, and seek a rigidly rotating configuration of stars
and standing waves.  (This idea goes back 30 years to Thorne and
nonspherical modes of neutron stars.)  If the cavity is too big, then
the total mass-energy of the standing waves dominates the problem  ---
in the limit of an infinite cavity, we get a non-asymptotically flat
spacetime.  However, thanks to GW inefficiency, the cavity can be
made rather large.  The best approach of this kind seems that of
Blackburn and Detweiler ({\sl Phys.\ Rev.\ }{\bf D46}, p.2318, 1992).
When applied to the ISCO radius, though, it gives a value which
is much lower than that from any other method.

Another way to define the ISCO precisely is to go to some kind of
radiation-less approximation to full GR\@.  (The waves can then be painted
in later.)  The most ambitious version to date of such an approach is that
of Wilson \& Mathews ({\em Phys.\ Rev.\ Lett.\ }{\bf 75}, p.4161, 1995),
and  Wilson, Mathews, \& Marronetti ({\em Phys.\ Rev.\ }{\bf D54}, p.1317,
1996;
\htmladdnormallink{gr-qc/9601017}{http://xxx.lanl.gov/abs/gr-qc/9601017}),
who numerically construct a family of curved but non-radiating spacetimes
containing fully hydrodynamic neutron stars.  They find a rather large ISCO
radius --- in fact, large enough that the orbital angular momentum $J$ still
exceeds $M^2$, where $M$ is the total mass-energy of the binary system, so
that the stars are forbidden to plunge directly to a Kerr black hole!
Another important result comes out of their work:  The neutron stars
themselves may go radially unstable and begin collapsing to black holes,
before the ISCO is reached.  The neutron star binary problem may reduce
itself to the black hole binary problem!   If so, matter is removed from
the game, and cannot carry angular momentum to infinity, or help make gamma
ray bursts.  Again, however, it's not clear how good the approximation
is.  For further aspects see
\htmladdnormallink{gr-qc/9512009}{http://xxx.lanl.gov/abs/gr-qc/9512009},
\htmladdnormallink{gr-qc/9601019}{http://xxx.lanl.gov/abs/gr-qc/9601019},
\htmladdnormallink{gr-qc/9603043}{http://xxx.lanl.gov/abs/gr-qc/9603043},
\htmladdnormallink{gr-qc/9701033}{http://xxx.lanl.gov/abs/gr-qc/9701033}.

As should be clear, a number of resourceful groups have studied this
problem by an amazing variety of methods, none rigorous to date --- and
it is sometimes hard to inter-compare results.  Here is a suggestion:
Everyone at work on this problem will benefit if all groups report,
at a minimum, the invariant observables $M$, $J$, and $\Omega$ (noting
that quadrupole gravity waves will show up at $2\Omega$), for each
orbital configuration, and especially for the ISCO:
\medskip
\begin{center}
\begin{tabular}{|l|c|c|}
\hline
{\bf Invariant Quantity} & {\bf Dimensional} & {\bf Normalized} \\
\hline
Binding Energy		& $M-M_\infty$ & $(M-M_\infty)/M_\infty$ \\
Orbital Angular Velocity & $\Omega$ & $\Omega M_\infty$ \\
Total Angular Momentum	& $J$	    & $J/M_\infty^2$ \\
\hline
\end{tabular}
\end{center}

\medskip

For a given binary, the evolutionary sequence of orbital configurations
{\em should} obey the Thorne-Zel'dovich law, $dM = \Omega dJ$\@.  I say
{\em should} rather than {\em does}, because to my knowledge no-one has
proved this law for configurations which are not strictly stationary, rigidly
rotating, asymptotically flat solutions of the full Einstein equations ---
so binaries are not yet covered, and in particular, radiation-less schemes
are not yet covered.  This provides a good sanity check on the numerical
calculations, assuming validity.  To prove more general validity poses a
good research problem.

\medskip

In summary, here is a list of important questions that remain open:
\begin{enumerate}
\item Do real neutron stars {\em disrupt}, or {\em collapse to black
holes}, or {\em neither}, before the ISCO?
\item How much matter (neutrinos, electromagnetic fields, relativistic
jet or wind, excretion disk) gets left behind from a coalescing neutron
star binary?
\item Do real (say, initially non-rotating) black holes reach the ISCO
while the total $J$ is too big for any Kerr black hole, $J>M^2$?  ---
If so, do unexpected GW signals emerge after the ISCO to carry off the
excess $J$?
\item Same question for real neutron stars.  In this case, does matter
or gravity waves carry off most of the excess $J$?
\end{enumerate}

\vfill\pagebreak
\section*{\centerline{Journ\'ees Relativistes '96}}
\addtocontents{toc}{\protect\bigskip}
\addtocontents{toc}{\bf Conference reports:}
\addtocontents{toc}{\protect\medskip}
\addcontentsline{toc}{subsubsection}{\it 
Journ\'ees Relativistes '96, by D. Brill, M. Heusler and 
G. Lavrelashvili}
\begin{center}
\medskip
Dieter Brill$^1$, Markus Heusler$^2$ and George Lavrelashvili$^3$\\

\medskip

1. University of Maryland, 
\htmladdnormallink{brill@umdhep.umd.edu}{mailto:brill@umdhep.umd.edu}\\
\medskip
2. ITP Zurich, Switzerland, 
\htmladdnormallink{heusler@physik.unizh.ch}{mailto:heusler@physik.unizh.ch}\\
\medskip
3. ITP Bern, Switzerland,
\htmladdnormallink{lavrela@butp.unibe.ch}
{mailto:lavrela@butp.unibe.ch}
\end{center}

The Journ\'ees Relativistes were started in the 1950's 
as an annual meeting of French-speaking relativists and 
cosmologists. In recent years they acquired
a more international flavor, visited various countries in Europe, 
and adopted English as their lingua franca. 
The general purpose of these meetings is to
report on progress in general relativity and its applications,
and to strengthen international scientific cooperation.
The major issues of the 1996 conference included 
general relativity, new developments in cosmology (driven by recent
observational results of the CMB and by gravitational lensing)
numerical relativity, quantum cosmology and quantum gravity.

The local Organizing Committee of the
XXVIth meeting consisted of Ruth Durrer (Geneva), 
Petr H\'{a}j\'{\i}\v{c}ek (Bern),
Philippe Jetzer (Zurich), George Lavrelashvili (Zurich),
Mairi Sakellariadou (Geneva), and
Norbert Straumann (Zurich), Chairman. 

The meeting could not have taken place without 
the financial support of
the ETH Zurich, the Swiss National Science Foundation, 
the Dr. Tomalla Foundation, and the  Hoch\-schulstiftung
of the University of Zurich.
 
The Journ\'ees Relativistes '96 were held May 25-30 
at the Centro Stefano Franscini on Monte Verit\`{a}
in Ascona, Switzerland, 
a center for small conferences providing accommodations, lecture 
hall, discussion rooms, meals, and spectacular views of Lago Maggiore.
The seventy-five participants came from fifteen 
countries, representing all
major centers of research in Europe, as well as representatives from
South Africa, Israel, Georgia, Russia, Mexico, Canada and USA. 
Extensive interaction between the participants was fostered by the
scheduling, which included adequate discussion time, and the
accommodations that allowed everyone to be in the same location;
even during evening excursions into the village colleagues were not
hard to find.

The main lectures were held by
B. Carr (London), P. Chrusciel (Tours), T. Damour (Bures-sur-Yvette),
S. Deser (Brandeis U), R. Durrer (Geneva), J. Fr\"ohlich (Zurich),
M. Heusler (Zurich), O. Lahav (Cambridge), A. Lasenby (Cambridge),
V. Mukhanov (Zurich), G. Neugebauer (Jena), A. Rendall (Potsdam),
C. Rovelli (Pittsburgh), P. Schneider (Garching), E. Seidel (Potsdam)
and G. Veneziano (CERN). 
In addition, over forty shorter 
contributions were given in two parallel sessions.
The lectures and contributions have been published in 
Helvetica Physica Acta, Vol. 69 (1996), Nos~3~\&~4.

The Journ\'ees also included a visit to a nearby solar observatory, and
one afternoon was reserved for an excursion and banquet.
This involved a variety of activities hard to find in one place except in
Switzerland: boat rides, a sub-tropical island, a hike in the mountains,
and dining al fresco. 

The XXVIth Journ\'ees continued the tradition of a very
high scientific and cultural level, and we expect 
many further, interesting meetings
to come in this series.

\vfill\pagebreak
\section*{\centerline{TAMA Workshop}}
\addtocontents{toc}{\protect\medskip}
\addcontentsline{toc}{subsubsection}{\it 
TAMA Workshop, by Peter Saulson}
\begin{center}
\medskip
Peter Saulson\\Syracuse University\\
\htmladdnormallink{saulson@suhep.phy.syr.edu}{mailto:
saulson@suhep.phy.syr.edu}
\end{center}

The TAMA Workshop on Gravitational Wave Detection
November 12-15, 1996, Saitama, Japan

   Gravitational wave detection specialists from around the world braved the
fearsome Tokyo evening rush hour on Monday, November 11, to reach the
idyllic National Women's Education Center at Musashi-Ranzan for the first
TAMA Workshop. The meeting was planned as a coming out party for
TAMA300, Japan's debutante in the present season of large interferometer
construction. Although rather petite for its class (the "300" refers to the 
instrument's 300 meter arms), its many charms were well displayed to
those in attendance. And before the meeting had concluded, there was 
specific discussion of plans for a strapping 3 km sister instrument hoped 
to follow in a few years.

  Talks at the meeting focused on progress around the world in technologies
necessary for successful detection of the feeble waves carrying across
cosmic distances the messages of the violent deaths of stars or compact
binary systems, or perhaps the birth cries of black holes. While proceeding
rapidly, interferometer design will have to hurry to keep up with the 
breakneck pace that the large construction efforts have achieved. Notably, 
at this meeting the reports from all of the world's approved large projects 
(LIGO, VIRGO, GEO 600, and TAMA300) included photographs of large quantities 
of concrete, in some cases still wet but in all cases demonstrating the
fruition of plans many years in the making. A glance back at this 
decades-long history would have been enough to bring tears to one's eyes, 
were it not for the bracing beauty of the promise of the years ahead.

  A sign of the pressing urgency of dreams come true is the attention
now being paid to ways to record, store, retrieve, and analyze the
mass quantities of data soon to be generated by the world's complement
of interferometers. Talks among the various parties, most recently and
intensively by LIGO and VIRGO, have led to the outlines of a specification
for a common format to be used for recording data; at the TAMA workshop the
Japanese scientists signaled their intention to aid in its design. This
development was singled out by LIGO P.I. Barry Barish in his toast at
the conference banquet, as a mark of the cooperative spirit that has
marked this international endeavor and as a crucial step to ensure
the linkage of interferometers into a worldwide observatory.

   The foreign visitors enjoyed the chance to become better acquainted with
the staff of the Japanese collaboration, especially with the students and
other young scientist who carry so much of the burden in large projects
such as this. A highlight of the meeting was the tour of the site
of TAMA300, snugly ensconced underground on the attractive campus of the
National Astronomical Observatory at Mitaka, in the Tokyo suburbs. All three 
buildings and their connecting 300 meter tunnels are complete, and large
vacuum hardware is everywhere in evidence. In the vertex building, visitors
had the chance to inspect the intriguing X-pendulum low frequency vibration
isolation system, under the proud and watchful eye of designer Mark Barton.
If the rest of the interferometer can be constructed as nicely as the parts
completed to date, then successful attainment of the design sensitivity
(rms strain of $3\times10^{-21}$) 
should be possible by the target date of 1999.

   Thus, TAMA300 should inaugurate the large interferometer era of
gravitational wave detection. If all goes according to plan, it will
be joined on the air by GEO 600 (near Hannover) in 2000. VIRGO plans to 
complete the construction of its 3 km interferometer in Cascina (near Pisa)
in 2001. LIGO will have completed construction of its two 4 km long 
interferometer sites in 1999, followed by a two year period of 
commissioning. By 2002, its three interferometers (one at Livingston, 
Louisiana and the 4 km / 2 km pair at Hanford, Washington) are expected to 
be on the air for an inaugural two year data run. In the meanwhile, the 
Japanese hope to begin work on a 3 km interferometer by 2000, filling out 
a network that truly spans the globe.

\vfill\pagebreak
\section*{\centerline{Midwest gravity meeting}}
\addtocontents{toc}{\protect\medskip}
\addcontentsline{toc}{subsubsection}{\it 
Midwest gravity meeting, by Comer Duncan}
\begin{center}
\medskip
Comer Duncan\\Bowling Green State University\\
\htmladdnormallink{gcd@chandra.bgsu.edu}{mailto:
gcd@chandra.bgsu.edu}
\end{center}

The sixth annual Midwest Relativity Conference was held at Bowling
Green State University in Bowling Green, Ohio on Nov. 1-2, 1996.
There were about 50 participants, most of whom gave talks.  Each talk
was 15 minutes long.  For the most part the topics covered fell into
the following categories: 1) Numerical Relativity, 2) Mathematical
Relativity, 3) Observational Topics, 4) Quantum Gravity, and 5)
Alternative Theories of Relativity.

{\em Numerical Relativity.} There were quite a few talks which utilize
numerical techniques, but several were more or less devoted to issues
which center on the numerical implementation of Einstein's theory in
two and three spatial dimensions, with and without matter.  Beverly
Berger discussed evolutions of $U(1)$ symmetric cosmologies, showing
the tracking of regions of large curvature using her $U(1)$ symmetric
code.  David Garfinkle talked about a new symplectic algorithm for the
evolution of mixmaster spacetimes which accurately evolves to late
times.  Jorge Pullin gave an overview of the successful 'close
approximation' for colliding black holes and discussed its salient
features and applications. Serge Droz discussed a numerical
investigation of black hole interiors. Keith Lockitch reported on his
investigations of how to numerically construct a family of
asymptotically flat initial data sets which are axisymmetric and
geodesically complete as counter-examples to the cosmic censorship
conjecture.  Comer Duncan gave a talk about a new hyperbolic solver
for vacuum axisymmetric spacetimes which combines the Geroch manifold
of trajectories approach with a symmetric hyperbolic form of
Einstein's equations, and discussed a numerical implementation and
tests on weak gravitational waves.  Mark Miller reported his
investigations into the utility of Regge Calculus for numerical
relativity, focussing on issues of consistency and stability. Grant
Mathews discussed his work on the appearance of instabilities in close
neutron star binaries.

{\em Mathematical Relativity.} There were a wide variety of topics in
mathematical relativity.  Ulrich Gerlach discussed some recent work on
a variety of topics, including paired accelerated frames, diffractive
scattering, and achronal spin.  Jean Krisch reported on some work on
string fluid stress energy.  Steve Leonard talked about the appearance
of tail effects for various types of fields in curved spacetimes. Ted
Quinn reported his investigations into an axiomatic approach to
electromagnetic and gravitational radiation reaction.  Shyan-Ming
Perng talked about his research on conserved quantities at spatial
infinity.  Ed Glass discussed the calculation of the Bondi mass from
Taub numbers, while Jim Chan discussed radiative falloff in
non-asymptotically flat spacetimes.  Juan Perez-Mercader presented his
views on how self-organized criticality may be expressed in the
universe.  Kevin Chan talked about modifications of the spinning BTZ
black hole when the theory contains a dilaton field.  Masafumi Seriu
gave an overview of his work to characterize the global geometrical
structure of a space in terms of the spectra of a suitable operator,
including discussion of a measure of closeness of two universes based
on the spectra. There were two talks utilizing torsion. Richard
Hammond discussed gravitation with torsion derived as the exterior
derivative of a second potential, built so that electromagnetism may
be generalized such that the source of the torsion gives rise to the
magnetic dipole moment of the electron.  Harry Ringermacher gave a
treatment in which gravity with torsion is present to encode the
electromagnetic field and discussed several solutions.

{\em Observational Topics.} Mark Beilby gave a talk about methods of
predicting test mass thermal noise by measurement of the anelastic
aftereffect in the context of the LIGO project.  Eric Poisson talked
about the dual use of space-based interferometers to measure black
hole parameters and as a means of testing general relativity. Andrew
de Laix gave a treatment of the gravitational lensing signature of
long cosmic strings.  Bob Wald gave an overview of some recent work on
galactic and smaller scale gravitational lensing, while Daniel Holz
followed with further developments and some applications.  Philip
Hughes discussed some recent high resolution relativistic
extra-galactic jet simulations as a possible probe of aspects of
super-massive black holes.

{\em Quantum Gravity.} The variety of topics which are quantum related
was quite wide in scope.  Leonard Parker gave a talk about his
research on 2D black holes, reporting on his work on the formation and
evaporation of 2D black holes in dilatonic quantum gravity. Louis
Witten talked about the formation and evaporation of naked
singularities in 2D which suggest that the naked singularity will not
exist but instead there would be a large outburst of radiation.  Ivan
Booth reported on the cosmological production of charged and rotating
black hole pairs.  James Geddes gave a discussion of some recent work
on whether there exists a measure on the space of all paths in
Schroedinger quantum mechanics such that the time evolution of the
system is given by an appropriate path integral. He give an example of
a system for which the answer is no.  Hong Liu talked about quantum
hair, instantons, and black hole thermodynamics.  Bob Mann then gave a
report on the pair production of topological anti de Sitter black
holes.  The talks by Michael Pfenning and Matt Visser were of a
different orientation.  Pfenning discussed quantum inequalities in
static curved spacetimes.  Visser gave an overview of the violation of
energy conditions at order hbar, showing how the polarization of the
vacuum by the semi-classical gravitational field causes a shift in the
stress-energy which violates all the classical energy conditions.
Rhett Herman talked about the use of the DeWitt-Schwinger
point-splitting technique to construct the stress-energy of a complex
scalar field in curved spacetime.

{\em Alternative Theories of Relativity.} Ken Seto discussed a special
relativity alternative.  Edward Schaefer talked about means of
eliminating black holes from relativity theory.

The Sixth Midwest Relativity Conference included a wide variety of
talks, giving ample evidence of the breadth of interests of
relativists in and around the midwest.  The meeting at Bowling Green
demonstrated that the midwest meetings have achieved a stable status.

Thanks to all who attended! The next meeting will be at Washington
U. in St. Louis with Wa Mo Suen the prime contact.  See you all at the next meeting!

\vfill\pagebreak
\section*{\centerline{OMNI-1 Workshop}}
\addtocontents{toc}{\protect\medskip}
\addcontentsline{toc}{subsubsection}{\it 
OMNI-1 Workshop, by N.S. Magalh\~aes, W.F. Velloso Jr and O.D. Aguiar}
\begin{center}
{\bf and the beginning of the International Gravitational
Radiation Observatory}\\

\medskip

N.S. Magalh\~aes, W.F. Velloso Jr and O.D. Aguiar\\
Divis\~{a}o de Astrof\'{\i}sica - INPE, S\~{a}o Paulo, Brazil\\
\htmladdnormallink{nadja@uspif1.if.usp.br}
{mailto:nadja@uspif1.if.usp.br},
\htmladdnormallink{velloso@das.inpe.br}
{mailto:velloso@das.inpe.br},
\htmladdnormallink{odylio@das.inpe.br}
{mailto:odylio@das.inpe.br},
\end{center}

The First International Workshop for an Omnidirectional Gravitational
Radiation Observatory (OMNI-1) was held at S\~ao Jos\'e dos Campos,
Brazil, in May 23-30, 1996.  It joined up some of the most active
scientists now working in the field of gravitational wave detection,
representing six countries\footnote{Australia, Brazil, Italy, The
Netherlands, Russia and USA.} and the major experimental projects in
the world\footnote{LIGO, VIRGO, AIGO, NAUTILUS, ALLEGRO, AURIGA,
ALTAIR, ELSA, GRAIL, PERTH and EINSTEIN.}.  The meeting allowed
researchers and students to present the state of art of the projects
and to exchange experiences among their groups.  Also, during the
opening talks and the six sections of OMNI-1 the participants had the
opportunity to discuss about many important issues regarding
gravitational wave research.

Several theorists contributed with interesting works on super-string
theory, cosmic strings, gravitational emission and theoretical
gravity. Also, talks on the emission of gravitational waves from
supernovae, black-holes collisions, binary coalescence and radio
pulsars were presented.

Experimentalists from both resonant-mass and interferometric projects
talked about virtually all the most important aspects concerning
gravitational wave experiments nowadays. For instance, the
detectability of gravitational signals generated by the coalescence of
binary systems, precessing neutron stars in the galaxy or core collapse
events was discussed, as well as the influence of cosmic rays on
resonant-mass detectors.  And several results were shown concerning
transducer construction techniques, thermal cooling methods and the
research on special materials to be used in spherical antennas.

Scientists representing the different projects talked about the
present status of their experiments and showed exciting new results
that suggest that more progress is expected in this field of research
for the next years. One of these breakthroughs is the construction of
a fourth generation of resonant-mass detectors using large spherical
antennas, which is under consideration by almost all the groups
working with this kind of detectors.

The concluding session of the Workshop (the roundtable section)
created an opportunity to increase the collaboration among the
groups. It was clear for the participants that in the field of
gravitational wave detection scientists are working at the edges of
technology and science so that international collaboration becomes
essential. Therefore they decided to formalize the OMEGA
Collaboration\footnote{ OMEGA homepage is on the WEB site
\htmladdnormallink{http://phwave.phys.lsu.edu/omega/}
{http://phwave.phys.lsu.edu/omega/}},intended to create an
International Gravitational Wave Observatory, composed by a network of
resonant-mass (using both bars and spheres as antennas) and
interferometric detectors , under coordinated operation. Such
observatory is expected not only to detect gravitational wave signals,
but also to determine their intensities and polarizations, as well as
the directions of their astrophysical sources, in a large spectrum of
frequencies.

The final result of the Workshop - the OMEGA Collaboration -
represents a major effort of scientists working on gravitational wave
detection from all around the world. Their intention is to join their
capabilities, experiences, resources and ideas to create a
revolutionary scientific tool and develop a new means to study the
universe: Gravitational Astronomy.

\vfill\pagebreak
\section*{\centerline{Chandra symposium}}
\addtocontents{toc}{\protect\medskip}
\addcontentsline{toc}{subsubsection}{\it 
Chandra symposium, by Robert Wald}
\begin{center}
\medskip
Robert Wald\\University of Chicago\\
\htmladdnormallink{rmwa@midway.uchicago.edu}
{mailto:
rmwa@midway.uchicago.edu}
\end{center}

	A Symposium on ``Black Holes and Relativistic Stars'' was held
on the University of Chicago campus on the weekend of December 14 and
15, 1996. The Symposium was dedicated to the memory of S.
Chandrasekhar, who died in August, 1995 and had devoted much of the
last 30 years of his scientific career to research in these areas.
Although the Symposium was originally envisioned as a relatively
``small'' meeting, it was attended by over 500 registered
participants, about half of whom came from outside of Chicago. Many of
the participants stayed to attend the ``Texas Symposium'', which was
held in downtown Chicago during the following week.
	The Symposium consisted of twelve plenary talks, each one hour
in length. Valeria Ferrari opened the Symposium on Saturday morning
with a review of her work, done in collaboration with Chandrasekhar, on
perturbations of black holes and relativistic stars. John Friedman
reviewed work on rotating relativistic stars, including the
information to be gained from millisecond pulsars on neutron star
matter. Kip Thorne brought us up to date on the status of LIGO and
VIRGO and reviewed what we might learn from them and LISA about black
holes and neutron stars.
	In the Saturday afternoon session, Martin Rees reviewed the
observational evidence for black holes, including some new, strong
evidence for black holes at the centers of galaxies. Roger Penrose
presented some perspectives on cosmic censorship and singularities.
Saul Teukolsky described some of the ongoing research in numerical
relativity, aimed at analyzing the collisions of black holes.
	In the Sunday morning session, Werner Israel discussed his
work and that of others on the internal structure of black holes. I
reviewed the status of black hole thermodynamics, with an emphasis on
the apparent ``universality'' of the laws. Rafael Sorkin then
presented some of his ideas related to the statistical origin of the
laws of black hole thermodynamics.
	The Sunday afternoon session began with James Hartle
describing generalized quantum theory and how it might treat issues
associated with black hole evaporation. Stephen Hawking reviewed his
ideas on the loss of quantum coherence due to black holes and briefly
described a new calculation related to this process. The Symposium
concluded with a relatively non-technical talk by Edward Witten, which
reviewed the development of ideas in string theory and gave his
present perspectives on ``quantum and stringy geometry''.
	The proceedings of the Symposium will be published by the
University of Chicago Press, and should be available in early 1998.

\vfill\pagebreak
\section*{\centerline{Penn State Meeting}}
\addtocontents{toc}{\protect\medskip}
\addcontentsline{toc}{subsubsection}{\it 
Penn State Meeting, by Lee Smolin}
\begin{center}
\medskip
Lee Smolin\\Penn State\\
\htmladdnormallink{smolin@phys.psu.edu}
{mailto:
smolin@phys.psu.edu}
\end{center}

On November 8-10, at University Park, the Fourth Annual Penn State
Conference took place. This year's conference was titled ``New voices
in relativity and quantum gravity'' and featured plenary speakers that
were at the postdoctoral level in their careers. It also had short
oral contributions in the afternoons which were allotted in a
first-come, first-serve basis, much along the model of the Pacific
Coast and Midwest meetings.
 
The plenary speakers were Greg Cook (Cornell), Simonetta Frittelli
(Pittsburgh), Gabriela Gonz\'alez (MIT-LIGO), Juan Maldacena
(Rutgers), Hans-Peter Nollert (Penn State), Amanda Peet (Princeton),
Thomas Thiemann (Harvard).

Between them they discussed new developments in gravitational physics, on
both the quantum and classical side.  Gonz\'alez told us about the
status of the construction of the LIGo detectors, while Cook and
Nollert reviewed the status of calculations to model astrophysical
events such as colliding black holes that may provide signals for LIGO.
The important new developments in string theory, that have made it 
possible to compute the entropies of extermal and near extremal
black holes were the subjects of the talks by Maldecena and Peet.
Thomas Thiemann described his recent work in non-perturbative quantum
gravity which results in the construction of at least one version of
quantum general relativity, while Frittelli described the work of
herself and her collaborators which leads to a reformulation of
general relativity in terms of the dynamics of null surfaces.

There were also a large number (41!) of contributed talks, spanning
most areas of interest in relativity.

Part of the idea of the meeting was to launch in the East a series of
meetings of similar spirit to that of the Midwest and Pacific Coast
meetings. The next such conference will be organized in Syracuse
University next year.

\vfill\pagebreak
\section*{\centerline{Aspen  Winter Conference}}
\addtocontents{toc}{\protect\medskip}
\addcontentsline{toc}{subsubsection}{\it 
Aspen Winter Conference, by Syd Meshkov}
\begin{center}
\vskip -.5cm
{\Large \bf on gravitational waves and their detection}\\
\medskip
Syd Meshkov\\Caltech - LIGO\\
\htmladdnormallink{syd@ligo.caltech.edu}
{mailto:
syd@ligo.caltech.edu}
\end{center}

The 1997 Aspen Winter Conference on Gravitational Waves and their
Detection (the third in this series) took place in Aspen, CO, January
26 to February 1st, 1997. It was originally planned to stress three
main areas, Advanced Detector Research and Development, Collaboration
Formation for Advanced Detectors, and the LIGO Research Community. By
the time that the meeting finished, an air of excitement and
exuberence prevailed. New and exciting ideas, both experimental and
theoretical had been introduced, and a number of collaborative efforts
had been organized. In addition, issues of a political and
organizational nature were raised and discussed openly.

The meeting was held under the auspices of the Aspen Center for
Physics. The morning sessions were held in the Flug Forum, a beautiful
auditorium in the new building of the Physics Center. The evening
sessions were held in the Conference Center of the Aspen Institute of
Humanistic Studies. Most of the participants were comfortably housed in
the nicely appointed apartments of the Institute. There were 61
participants, including representatives of all existing and planned
interferometers and bars.

Reports on the status of Virgo, GEO 600, LIGO , ACIGA, TAMA 300, and
Bar Detectors by Giazotto, Hough, Shoemaker, Munch, Barton and Hamilton
framed the conference.

Because the time has come for serious discussion of how to best
organize the physics effort that will operate and do research with
LIGO, several sessions were devoted to gathering the views of the
participants, working towards a consensus, and taking early steps in
the formation of the requisite Collaboration. Discussions of how to
initiate formation of the LIGO Collaboration started early in the
Conference in a session on Monday night, following introductory remarks
by Gary Sanders. These discussions continued throughout the Conference,
both formally and informally. They were addressed in the LIGO Research
Community formal session, where Dave Berley presented the NSF views of
LIGO structure, management, and goals. A recapitulation of the
discussions was given in the final session by Finn and Saulson.

With construction of the LIGO Observatories well underway, and
following the recommendations of the NSF McDaniel Report, the 1997
Conference focused on Research and Development for the next generation
of detectors. Imaginative ideas for advanced interferometry,
suspensions and seismic isolation, lasers, and optics were presented
and freely discussed in both formal presentations and in many informal
meetings. The GEO, JILA, Stanford and Virgo groups participated
actively in these discussions.

In a theoretical vein, Jim Wilson and Grant Mathews presented work on
General Relativistic Numerical Hydrodynamics for Neutron Star Binaries
which questions the utility of the templates which are currently being
implemented for analyzing neutron star binaries. Naturally, this
engendered much discussion, and was addressed in Alan Wiseman's
presentation. Bruce Allen and Sam Finn engaged in a colloquy on how to
analyze data, and the bar people described how they extract signals
from their data.

Tuck Stebbins reported on a possible speeded up LISA system, to be launched in
the year 2004, instead of in 2015, that uses NASA vehicles together with
existing LISA technology. Choptuik showed where the Grand Challenge
stands in its final year of funding.

The conference concluded with summaries by Finn and Saulson.  The
general feeling was that communications had been greatly enhanced
between groups with different agendas. This was especially so in the
"neutral" location of beautiful Aspen. The formation of the LIGO
collaboration has been greatly facilitated.  One important outcome of
the conference was the formation of two working groups, one on high
power laser R\&D, the other on suspensions and seismic isolation.
Exciting new physics questions have been raised, and the imaginative
proposals for detection techniques stimulated everyone.

As for the past two conferences, a volume of all of the transparencies
shown at the conference will be distributed to each participant shortly
after the conference.

\end{document}